\documentclass[11pt]{article}
\usepackage{blois,epsfig}

\def\be{\begin{equation}}
\def\ee{\end{equation}}
\def\bea{\begin{eqnarray}}
\def\eea{\end{eqnarray}}

\begin{document}

\title{ELASTIC MESON-NUCLEON AND NUCLEON-NUCLEON SCATTERING:\\
MODELS vs. ALL AVAILABLE DATA \footnote{Contribution to the
proceedings of the XIth International Conference on Elastic and
Diffractive Scattering Ch\^{a}teau de Blois, France, May 15 - 20,
2005 }}

\author{ E. MARTYNOV $^{\dagger}$, J.R. CUDELL $^{\ddag}$ and A. LENGYEL $^{\sharp}$ }

\address{$\dagger$ Bogolyubov Institute for Theoretical Physics, Kiev, UA-03143, Ukraine,\\
$^\ddag$ Physique th\'eorique fondamentale, D\'ep. de Physique,
Universit\'e de  Li\`ege,
B-4000 Li\`ege~1, Belgium,\\
$^{\sharp}$ Institute of Electron Physics, Universitetska 21,
UA-88000 Uzhgorod, Ukraine. }

\maketitle\abstracts{ We consider simple-pole descriptions of soft
elastic scattering at small $t$ and $s$, and allow for the presence
of a hard pomeron. We have built and analyzed an exhaustive dataset
and we show that simple poles provide an excellent description of
the data in the region $- 0.5$ GeV$^2 < t <$ $- 0.1$ GeV$^2$,
6~GeV$<\sqrt{s}<$63~GeV. We show that new form factors have to be
used, and get information on the trajectories of the soft and hard
pomerons.}

Recently \cite{clms}, we have shown that a model that includes a
hard pomeron reproduces very well (with $\chi^{2}/d.o.f.\approx
0.95$ or with confidence level CL=93\%) total cross sections and the
ratio $\rho$ of the real to imaginary parts of the forward
scattering amplitude, while the description obtained from a soft
pomeron is much less convincing \cite{COMPETE}
($\chi^{2}/d.o.f.\approx 1.07$, CL=6\%). We considered the full,
standard set of forward data \cite{t0set} for $pp$, $\bar p p$,
$Kp$, $\pi p$, $\gamma p$ and $\gamma\gamma$, and showed that the
description extends down to $\sqrt{s}=5$ GeV.

The details of these results are presented in the contribution of
J.R. Cudell to these proceedings \cite{JRCblois}. However, despite
an excellent $\chi^{2}$ and the fact that the hard pomeron intercept
is very close to what is observed in deeply inelastic scattering
\cite{DisL} and in photoproduction \cite{DppL}, it is not entirely
sure that it is present in soft scattering. Indeed, its couplings
are small and its contribution is less than 10\% for
$\sqrt{s}<100$~GeV. Hence it is important to look for confirmation
of its presence in other soft processes, and the obvious place to
start from is elastic scattering. This analysis is briefly presented
in this paper.


{\bf 1. Scattering amplitudes.} We parametrise all exchanges by
simple poles, and limit ourselves to a region in $s$ and $t$ where
these are dominant. The normalization of the amplitude $A^{ab}(s,t)$
that describes the elastic scattering of hadrons $a$ and $b$ is
given by
\begin{equation}
\label{eq:sigtot} \sigma_{tot}^{ab}(s)=\frac{1}{2q_{ab}\sqrt{s}}\Im
mA^{ab}(s,0), \qquad
\frac{d\sigma_{el}^{ab}(s,t)}{dt}=\frac{1}{64\pi
sq_{ab}^{2}}|A^{ab}(s,t)|^{2},
\end{equation}
where
$q_{ab}=\sqrt{[s_{ab}^{2}-4m_{a}^{2}m_{b}^{2}]/4s}\,\,(s_{ab}=s-m_{a}^{2}-m_{b}^{2})$
is the momentum of particles $a$ and $b$ in the center-of-mass
system.

Regge theory implies that one can write $A(s,t)\equiv A(z_{t},t)$,
where the Regge variable,
$z_{t}=(t+2s_{ab})/\sqrt{(4m_{a}^{2}-t)(4m_{b}^{2}-t)}$ is the
cosine of the scattering angle in the crossed channel. Absorbing in
the $t$-dependent factors the coupling functions of the standard
contribution of a Regge pole, one can write for the case of
scattering of $a$ on protons
\begin{equation}
A_{R}^{ap}(\tilde s_{ap},t)=\eta_{C}g^{a}_{R}F^a_R(t)F^{p}_{R}(t)\
 (-i\tilde
s_{ap})^{\alpha_{R}(t)}. \label{eq:pole}
\end{equation}
with $F_a^R(0)=1$, $a=$ $p$, $\pi$, $K$ and where $\tilde
s_{ab}=(t+2s_{ab})/s_{0}, \quad  s_{0}=1$ GeV$^{2}$. For a
crossing-even, $C=+1$, (resp. crossing-odd, $C=-1$) reggeon
$\eta_{C}=-1 \,(\mbox{resp.}\, i)$.

The model that we are considering can be written:
\begin{equation}
A^{ap}(s,t)=A_+^{ap}(\tilde s_{ab},t)+A_S^{ap}(\tilde
s_{ab},t)+A_H^{ap}(\tilde s_{ab},t)\mp A^{ap}_-(\tilde s_{ap})
\label{eq:amplitude}
\end{equation}
with the $-$ sign for the (positively charged) particles.

{\bf 2. The data.} Most of the measurements, made throughout the
last 40 years, have been communicated to the HEPDATA group
\cite{HEPDATA}, so that one does not need to re-encode all the data.
However, some basic work still needs to be done, as there are 80
papers, with different conventions, and various units. The global
dataset contains 10188 points (at $t\neq 0$). Our detailed analysis
of these data and their systematic errors, list of the subsets
(measurements at different energies and momentum transfers) will be
given in a forthcoming paper \cite{clm2}. The statistics of the data
used in the present analysis is given in Table~\ref{tab:stats}. A
few subsets (5\% of whole set) were eventually excluded from the
fits because they strongly contradict \cite{clm2} the main part of
the data (see section 5).

\begin{table}[h]
\begin{center}
\caption{The statistics of the full dataset and of the present
analysis.} \label{tab:stats}
 \small {
\begin{tabular}{|c|c|c|c|c|c|c|c|} \hline
observable&$N_{pp}$&$N_{\bar p p}$&$N_{\pi^+ p}$&$N_{\pi^- p}$&$N_{K^+ p}$&$N_{K^-p}$&$N_{tot}$\\
\hline
 $d\sigma_{el}/dt$ (full set)     &4639& 1252& 802& 2169& 595& 731&10188\\
          this analysis&818 &281  &290 &483  &166 &169&2207\\
          after exclusion&795 &226  &281 &478  &166 &169&2115\\
\hline
\end{tabular}}
\end{center}
\end{table}

{\bf 3. Local fits.} In order to obtain the possible form factors,
we scan the dataset at fixed $t$, {\it i.e.} we fit small windows in
$t$ to a complex amplitude with constant form factors (and refer to
these fits as $local$ fits). The constants that we get will then
depend on $t$ and give us a picture of the form factor. The value of
the $\chi^2$ will also tell us in which region of $t$ we should
work.

This strategy however will not work for the general case considered
here: each bin does not contain enough points to have a unique
minimum. We can take advantage of the fact that both models
considered here give compatable intercepts for the crossing-odd and
crossing-even reggeon contributions. We can also read off the slopes
from a Chew-Frautschi plot. This gives the following $f/a_2$ and
$\rho/\omega$ trajectories:
\begin{equation}
\alpha_+= 0.61+0.82~t, \qquad  \alpha_-= 0.47+0.91~t \label{eq:Chew}
\end{equation}
Furthermore, we shall not be able to include a hard pomeron in the
local fits as its contribution is too small to be stable.

\medskip \noindent
\begin{minipage}{7.5cm}
We fit the data from 6 GeV$\leq\sqrt{s}\leq 63$ GeV, and we choose
small bins of width 0.02 GeV$^2$. We correct for the bin width by
assuming that within the bin the differential cross section can be
approximated by an exponential. We restrict ourselves to independent
bins where we have more than four points for each process.

\medskip
\includegraphics[scale=.4]{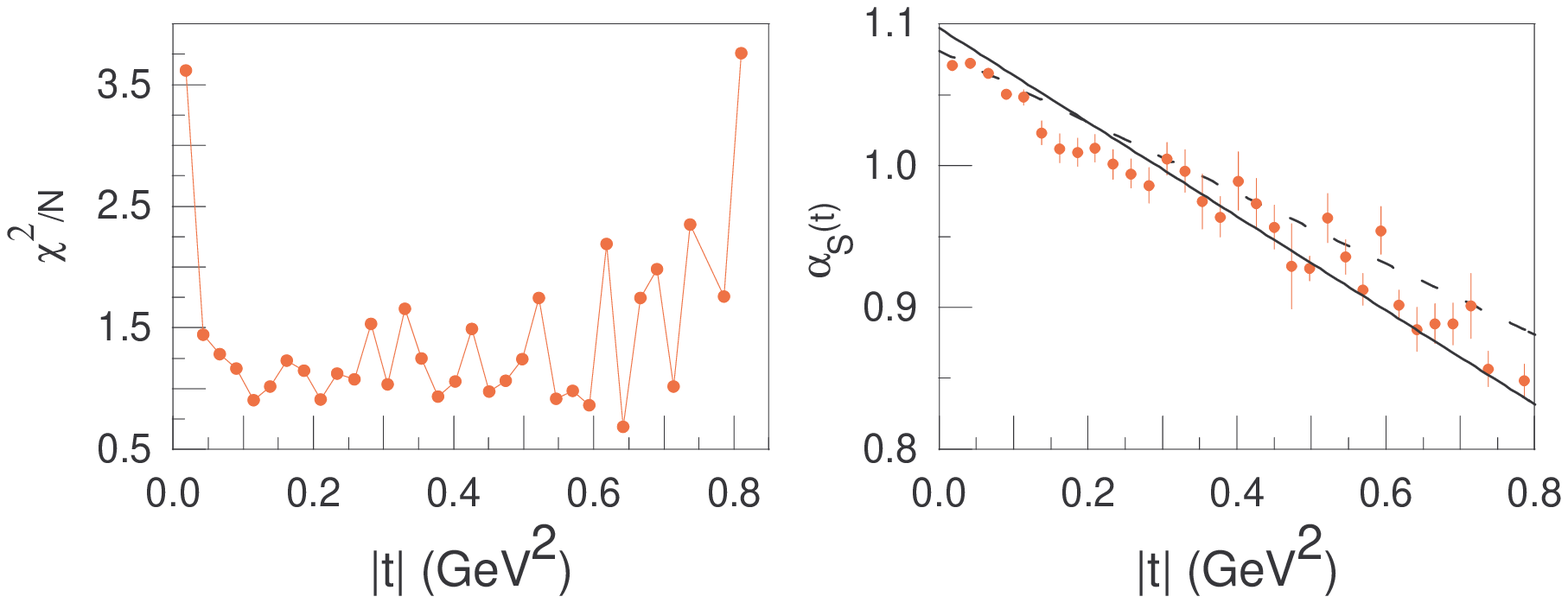}

\small{ { \bf Fig.1} The results of the local fits for the $\chi^2$
per number of points (left) and for the pomeron trajectory (right).
The dashed curve is from \cite{DLel} and the solid curve results
from the global fit given in the section 5.}
\end{minipage}
\hspace{0.5cm}
\begin{minipage}{7.5cm}
\begin{center}
\includegraphics[scale=.38]{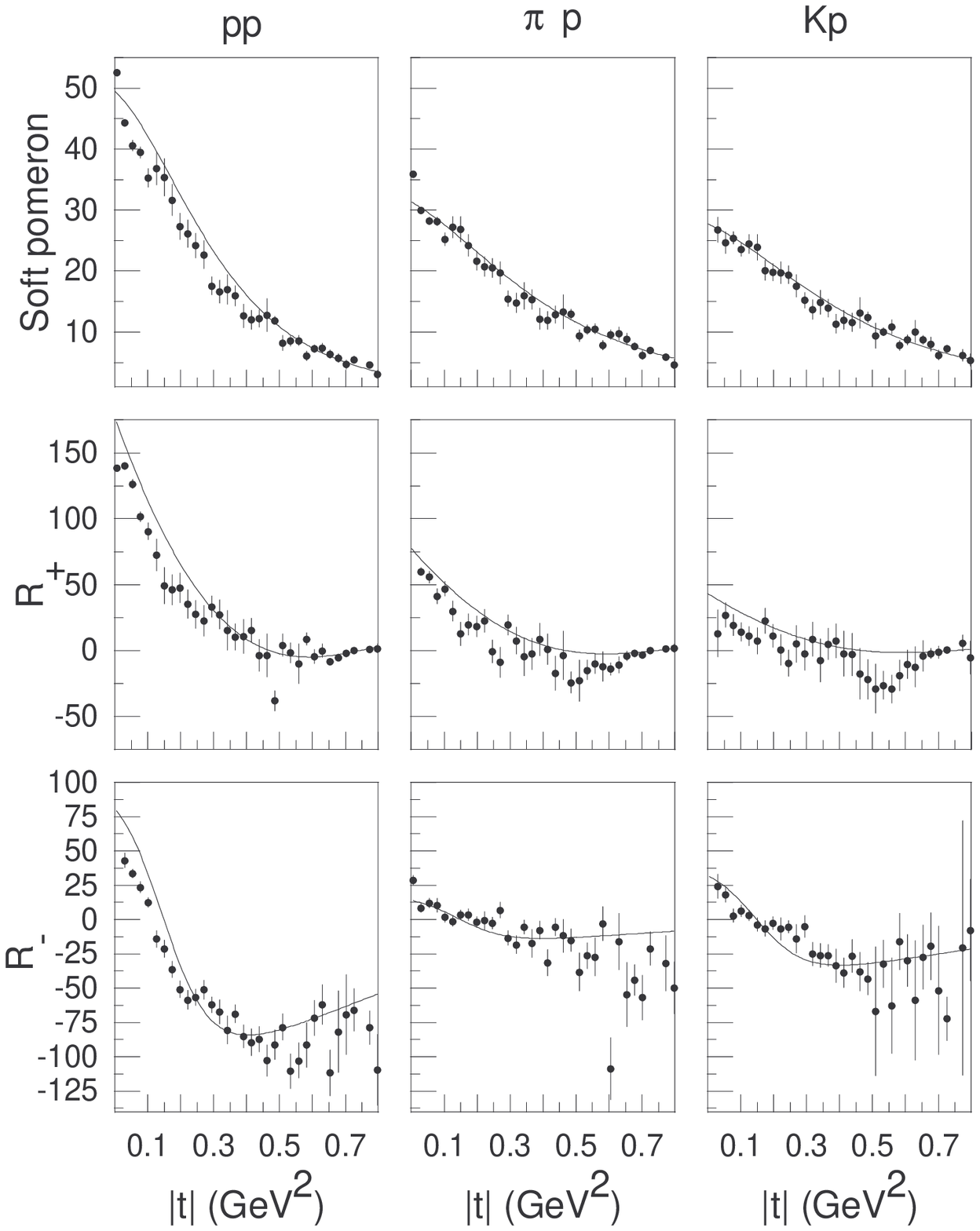}
\end{center}
\small{{ \bf Fig.2} The results of the local fits for the residues
of the poles. The curves are the results of a global fit explained
in the section 5.}
\end{minipage}
\medskip

Each of these fits gives us a values of the $\chi^2$ per number of
points, $\alpha_S(t)$ for each $t$, as well as the coefficients
$g_R^{ap}F^p_R(t)F^a_R(t)$. The results given in Figs.~1 and 2 show
two things: 1) the local fit to all data is never perfect, this can
be traced back to incompatibilities in the data\, \footnote{The
inclusion of data for $\sqrt{s}\leq 6$ GeV would only make this
problem worse.}; 2) the simple-pole description of the data has a
chance to succeed in a limited region: the $\chi^2$ grows fast both
at low $|t|$  and for $|t|>0.6$. To be conservative, we shall
consider a global fit in the region $0.1\leq |t|\leq 0.5$.

The right-hand graph in Fig.~1 shows the soft pomeron trajectory. It
is very linear as a function of $t$. Its intercept and slope are
somewhat different from the standard ones \cite{DLel}.

{\bf 4. Form factors.} Figure 2 shows the results for the residues
of the poles $g^a F_{R}^a(t)F_{R}^p(t)$. In all cases, it is obvious
that form factors must be different for different trajectories.

We find that  we can get a good description if we take ($a=\pi, K$)
\begin{equation}\label{eq:sffp2}
F^{p}_{S}(t)=\left
[1-t/t_{S}^{(1)}+\left(t/t_{S}^{(2)}\right)^2\right ]^{-1}, \,
F^{p}_{H}(t)=(1-t/t_{H})^{-2}, \,
F^{a}_{S}(t)=F^{a}_{H}(t)=(1-t/t^{a}_{S})^{-1}.
\end{equation}
In the $pp$ and $p\bar p$ cases, other dipoles are necessary to
describe the $C=\pm 1$ exchanges:
\begin{equation}\label{eq:fpik}
F^{p}_{C}(t)=(1-t/t_{C})^{-2}.
\end{equation}

In the crossing-odd case, the presence of a zero in form factors is
certain (see Fig.~2): this is the well-known cross-over phenomenon
\cite{crossovers}: the curves for $d\sigma/dt$ for $pa$ and $p\bar
a$ cross each other at some value of $t$. This phenomenon is present
at several energies, and seems to exist for all processes, for which
the cross-over point is at $|t| \approx 0.1 - 0.2$ GeV$^{2}$\,
\footnote{We tried obtaining these zeros through the rescattering of
various trajectories. However, their position and existence then
depend strongly on $s$.}. A similar zero (but less pronounced) is
seen for crossing-even form factors. Taking into account these facts
we  thus parametrise the $C=\pm 1$ reggeons ($a=p, \pi, K$)
\begin{equation}
A_{C}^{ap}(\tilde
s_{ap},t)=\eta_{C}Z^{a}_{C}(t)g^{a}_{C}F^a_{C}(t)F^{p}_{C}(t)\
(-i\tilde s_{ap})^{\alpha_{C}(t)}\quad \mbox{with}\quad
Z^{a}_{C}(t)= \frac{\tanh(1+t/\zeta_{C})}{\tanh(1)}.
\label{eq:pole-}
\end{equation}
Thus, the factor $Z^{a}_{C}(t)$ has a common zero $\zeta_{C}$,
independent of $s$ for $p,\pi,K$, but a different one for $C=+1$ and
$C=-1$ reggeons. The given form of $Z$ restricts its fast growth
with $t$.

{\bf 5. Global fits.}

\medskip \noindent
\begin{minipage}{8.5 cm}
Equipped with the information from the local fits, we can now
perform a global fit to the elastic data for 0.1 GeV$^2\leq |t|\leq$
0.5 GeV$^2$, for 6 GeV$\leq \sqrt{s}\leq $ 63 GeV. Fitting with the
full data set we obtained $\chi^{2}/d.o.f.\approx 1.45$ for the
model with only a soft pomeron (S) and $\approx 1.33$ for the model
with soft and hard pomerons (S+H). Excluding the subsets  which
contradict other data (only 92 points) we obtained the results shown
in the Table 2. The values of fitted parameters will be given in
\cite{clm2}. Here we give only the intercepts and slopes of pomeron
trajectories.
\end{minipage}
\hspace{0.4cm}
\begin{minipage}{6.5cm}
\begin{center}
\small {Table 2: Differential cross sections: partial values of
$\chi^{2}$.}
\end{center}
\begin{tabular}{|l|r||c||c|}
\hline Quantity     &  Number   & {S} & {S+H}         \\
\cline{3-4}
             &  of points  & $\chi^{2}/N$  & $\chi^{2}/N$                      \\
\hline \hline
$d\sigma^{pp}/dt$       &N=795 &  0.90 & 0.86  \\
$d\sigma^{\bar pp}/dt$  & 226 &  1.01 & 0.99  \\
$d\sigma^{\pi^{+}p}/dt$ & 281 &  0.90 & 0.89  \\
$d\sigma^{\pi^{-}p}/dt$ & 478 &  1.18 & 1.18  \\
$d\sigma^{K^{+}p}/dt$   & 166 &  1.02 & 1.11  \\
$d\sigma^{K^{-}p}/dt$   & 169 &  1.18 & 1.12  \\
\hline Totally          &2115 &  1.022& 0.997 \\
 \hline
\end{tabular}
\end{minipage}
\medskip

In the model with only a soft pomeron $\alpha_{S}(t)=1.0927+0.332\,
t$. If a hard pomeron is included then $\alpha_{S}(t)=1.0728+0.297\,
t$ and $\alpha_{H}(t)=1.45+0.1(\pm 0.2)t$. In both cases the slope
of the soft pomeron trajectory is higher than the standard one
\cite{DLel}.

{\bf Concluding remarks.} We have elaborated a complete dataset,
including an evaluation of the systematic errors for all data.

We showed that different reggeons must have different form factors.
We confirm that crossing-odd meson exchange has a zero. We also
found evidence for a sharp suppression of the crossing-even form
factor around $|t|=0.5$ GeV$^{2}$.

Because of the quality of the soft pomeron fit, the elastic data do
not confirm strongly the need for a hard pomeron. However it is
remarkable that the hard pomeron fit gives 0.1 GeV$^{-2}$ for the
central value of the slope, in agreement with \cite{DisL}.

Both models considered there lead to a very good fit that extends
well to S$p\bar p$S and Tevatron energies.

We hope that the complete dataset constructed here will serve as a
starting point for precise studies of the whole range of elastic
scattering and for the comparison of various models.

{\bf Acknowledgments.} E.M. thanks Prof. J.~Tran~Thanh~Van for the
support allowing to take part in the conference.

\end{document}